\documentclass[aps,pre,twocolumn,floatfix,superscriptaddress,nofootinbib,showkeys]{revtex4}
\usepackage{graphicx}

\newcommand {\be} {\begin{equation}}
\newcommand {\ee} {\end{equation}}
\newcommand {\Be}{\begin{eqnarray*}}
\newcommand {\Ee} {\end{eqnarray*}}
\newcommand {\bey} {\begin{eqnarray}}
\newcommand {\eey} {\end{eqnarray}}
\newcommand{\bit}{\begin{itemize}}
\newcommand{\eit}{\end{itemize}}
\newcommand{\bfl}{\begin{flusleft}}
\newcommand{\efl}{\end{flusleft}}
\newcommand{\bfr}{\begin{flushright}}
\newcommand{\bc}{\begin{center}}
\newcommand{\ec}{\end{center}}
\newcommand{\ben}{\begin{enumerate}}
\newcommand{\een}{\end{enumerate}}

\newcommand{\comment}[1]{}

\begin{document}

\title{Coherent response of the Hodgkin-Huxley neuron in the high-input regime}

\author{Alessandro Torcini}
\email{alessandro.torcini@isc.cnr.it}
\affiliation{Istituto dei Sistemi
Complessi  - CNR,
via Madonna del Piano 10, I-50019 Sesto Fiorentino, Italy}
\affiliation{INFN Sezione di Firenze,
via Sansone, 1 - I-50019 Sesto Fiorentino, Italy}
\author{Stefano Luccioli}
\affiliation{Istituto dei Sistemi
Complessi - CNR,
via Madonna del Piano 10, I-50019 Sesto Fiorentino, Italy}
\affiliation{INFN Sezione di Firenze,
via Sansone, 1 - I-50019 Sesto Fiorentino, Italy}
\author{Thomas Kreuz}
\affiliation{Istituto dei Sistemi Complessi - CNR,
via Madonna del Piano 10, I-50019 Sesto Fiorentino, Italy}

%%%%%%%%%%%%%%%%%%%%%%%%%%%%%%%%%%%%%%%%%%%%%%%%%%%%%%%%%%%%%%%%%%%%%%%%%%%
\begin{abstract}
We analyze the response of the Hodgkin-Huxley neuron to a large number of uncorrelated 
stochastic inhibitory and excitatory post-synaptic spike trains. 
In order to clarify the various mechanisms responsible for noise-induced 
spike triggering we examine the model in its silent regime.
We report the coexistence of two distinct coherence resonances:
the first one at low noise is due to the stimulation of 
{\it correlated} subthreshold oscillations; the second one at 
intermediate noise variances is instead related to the regularization 
of the emitted spike trains.
\end{abstract}
%%%%%%%%%%%%%%%%%%%%%%%%%%%%%%%%%%%%%%%%%%%%%%%%%%%%%%%%%%%%%%%%%%%%%%%%%%

%%%\pacs{05.45.-a,87.10.+e,87.17.Aa,05.40.Ca}
% 87.19.La - Neuroscience
% 84.35.+i Neural networks
% 05.45.-a Nonlinear dynamics and nonlinear dynamical systems
% 05.45.Xt Synchronization; coupled oscillators
% 87.10.+e General theory and mathematical aspects (biophysics)
% 07.05.Mh Neural networks, fuzzy logic, artificial intelligence
% 89.75.-k Complex systems
% 87.17.Nn Electrophysiology of nerve cells
% 87.17.Aa Theory and modeling; computer simulation (cellular process)
%05.40.Ca        Noise
%05.45.Tp        Time series analysis

\keywords{Hodgkin-Huxley model, Coherence, Noise, Sub- and Suprathreshold oscillations}

\maketitle

\subsection{Introduction}

In the last decades a large number of studies have been
devoted to the characterization of the response of simple and more
elaborated neuronal models under the influence of a large variety
of stochastic inputs~\cite{tuckwell,koch,ger}.
One of the main reasons justifying the interest
of neuroscientists for this subject resides in the observation
that {\it in vivo} neocortical neurons are subjected to a constant
bombardment of excitatory and inhibitory post-synaptic potentials
(EPSPs and IPSPs), somehow resembling a background noise \cite{destexhe}.

 Among the many proposed biophysical models the Hodgkin-Huxley model~\cite{hh} can still be considered
as a reasonably valid framework for exploring the main features of neural 
dynamics~\cite{hh_devil}. In order to understand the origin of the
variability observed in the distribution of spikes emitted by cortical neurons~\cite{shadlen_newsome_1998}
the response of the Hodgkin-Huxley (HH) model has recently been
studied under the influence of additive noise~\cite{tiesinga_2000,aguera03,rudolph_2003}
and subjected to trains of post-synaptic potentials~\cite{brown_feng_1999,hasegawa,tiesinga_2000,rudolph_2003}.

One of the most interesting phenomena observed experimentally
and numerically for excitable neuronal systems driven by noise is {\it coherence resonance} (CR)~\cite{gang,pikovsky_kurths_1997}:
it refers to the regularization of noise-induced oscillations occuring for
finite (not zero) noise intensity in the absence of any external injected signal
(for a comprehensive review see \cite{lindner_ojalvo_2004}).
Evidences of CR have been reported experimentally
for the cat's spinal and cortical neural ensembles \cite{manjarrez}
and theoretically for various neuronal models~\cite{lindner_ojalvo_2004}.
In particular, CR has been observed for the HH model~\cite{lee_neiman_1998,yu_wang_2001},
but these results mainly refer to additive continuous noise.

Our aim is to perform a detailed analysis of the response of the HH model
in the silent regime when it is subjected to many (hundreds or thousands) 
stochastic trains of post-synaptic potentials (PSPs) per emitted
spike (i.e. the system is in the so-called {\it high-input regime}~\cite{shadlen_newsome_1998,destexhe}).

The different mechanisms responsible for noise-induced neuronal firing are analyzed in
terms of statistical indicators (interspike-interval distributions and
their first moments) as well as of dynamical indicators (autocorrelation functions).
The analysis of the correlation time reveals a CR associated to the stimulations of 
coherent subthreshold oscillations around the rest state. This new type of CR coexists with 
{\it standard} CR related to the regularization of spike sequences~\cite{lindner_ojalvo_2004}.

\subsection{Model and methods}
\label{model}

The HH model describes the dynamical
evolution of the membrane potential $V(t)$. It can be written as
\bey
C\frac{dV}{dt}&=&g_{Na}m^{3}h(E_{Na}-V)+g_{K}n^{4}(E_{K}-V)+
\label{hh}
\\
&+&g_{L}(E_{L}-V) + I(t)\quad,
\nonumber
\eey
where $I(t)$ represents the synaptic current and the evolution of the
gating variables $X=m,n,h$ is ruled by three ordinary
differential equations of the form
$ dX/dt=\alpha_X(V)(1-X)-\beta_X(V)X $.
The parameters entering in Eq. (\ref{hh}) 
and the expressions of the nonlinear functions $\alpha_X(V)$ and
$\beta_X(V)$ are reported in \cite{noi}.

In this study we consider the response of the single HH model to $N_e$ (resp. $N_i$) uncorrelated 
trains of EPSPs (resp. IPSPs) with $N_e (N_i)$  ranging from 10 to 10,000.
By following Refs.~\cite{brown_feng_1999,feng_zhang_2001},
each PSP is schematized as an instantaneous
variation of the membrane potential by a positive (resp. negative)
amount $\Delta V$ for excitatory (resp. inhibitory) synapses.
The amplitude of each voltage kick is assumed to be 0.5 mV, i.e.
reasonably small ($\approx 7$ \%) with respect to the distance
between the "threshold" for spike initiation for rapid EPSPs
and the resting potential ($\approx 6 - 7 $ mV) estimated for the HH model~\cite{noble&stein,tuckwell}.
Moreover, amplitudes $\approx 0.5$ mV are comparable with average
EPSPs experimentally measured for pyramidal neurons in the visual cortex of rats~\cite{koch}.
This amounts to exciting the neuron (\ref{hh}) with an impulsive current
\begin{equation}
I(t)=Q\Big[\sum_{k=1}^{N_{e}}\sum_{l} \delta(t-t_{k}^{l})
-\sum_{m=1}^{N_{i}}\sum_{n} \delta(t-t_{m}^{n})\Big]
\label{I_imp}
\end{equation}
where $t_{k}^{l}$ (resp. $t_{m}^{n}$) are the arrival times of the
EPSPs (resp. IPSPs) and $Q=C \Delta V$ is the charge
associated to each kick. 
In order to mimic the inputs received by cortical
neurons, for each afferent (excitatory and inhibitory) synapse the time interval distribution between PSP
inputs is chosen to be Poissonian with an average frequency $\nu_0 = 100$ Hz~\cite{shadlen_newsome_1998}
\footnote{It should be stressed that the model here analyzed is driven by
impulsive post-synaptic currents and it should not be confused
with conductance-driven models~\cite{tiesinga_2000,rudolph_2003}.}.

The stochastic input can be characterized in terms of the average and the variance of the net
spike count within a temporal window $\Delta T$
\begin{equation}
N(\Delta T) = \sum_{k=1}^{N_e} n^E_k(\Delta T) - \sum_{m=1}^{N_i} n^I_m(\Delta T)
\label{sp_c}
\end{equation}
where $n^E_k(\Delta T)$ (resp. $n^I_m(\Delta T)$) represents the
number of afferent EPSPs (resp. IPSPs) received from neuron $k$ (resp. $m$)
in the interval $\Delta T$. In the high input regime, by assuming uncorrelated
input spike trains, $N(\Delta T)$ follows a Gaussian distribution (cf. \cite{noi}),
with average and variance derivable within the shot noise formalism as
\begin{equation}
< N(\Delta T) >=(N_e-N_i) \Delta T \nu_0 
\end{equation}
and
\begin{equation}
Var[N(\Delta T)]=(N_e+N_i) \Delta T \nu_0 = \sigma^2 \Delta T \nu_0.
\end{equation}
The parameter $\sigma=\sqrt{(N_e+N_i)}$ measures the standard deviation
of the stochastic input process.

The average current stimulating the neuron is given by
\begin{equation}
\bar I= \frac{C \Delta V <N(\Delta T)>}{\Delta T} = C \Delta V
\nu_0 (N_e - N_i).
\label{I_ave}
\end{equation}
$\bar I$ represents the bifurcation parameter ruling the dynamics of the deterministic 
HH model; for $\bar I < I_{SN} \simeq 6.27 \mu$A/cm$^2$ the model is in the {\it silent regime},
i.e., in the absence of noise the dynamics is attracted towards a stable fixed point 
and the neuron does not fire. However, since the fixed point is a focus
the relaxation towards the rest state occurs via damped 
oscillations ({\it subthreshold oscillations}) 
\cite{rinzel_miller_1980}. Periodic firing is observed for the HH model only above 
$I_{SN}$ and it is associated to frequencies in the range $50-80$ Hz.

We have verified that in the high input regime the response of the
neuron (for fixed $\nu_0$) is determined once fixed $\bar I$ and
$\sigma$, therefore it does not depend separately on $N_e$  and $N_i$,
but only on their difference $(N_e - N_i) \propto \bar I$ 
and sum $(N_e + N_i) \propto \sigma^2$~\cite{noi}. 

In order to characterize the output of the neuron and
to examine the coherence effects in the response we
have employed the following indicators:
the distribution of the output interspike intervals
$P_{ISI}(t)$ and its first moments; 
the coefficient of variation of the output interpike intervals distribution ($P_{ISI}(t)$) defined as
$R={S_{ISI}}/{A_{ISI}} $, where $A_{ISI}$ is the average 
and $S_{ISI}$ the standard deviation of the $P_{ISI}(t)$ (for a perfectly periodic response we have $R=0$ and $R=1$ for Poissonian output;
the normalized autocorrelation function $C(t)$ for the membrane potential and the
correlation time~\cite{pikovsky_kurths_1997} defined as
\begin{equation}
\tau_c= \int_0^\infty C^2(t) \enskip dt
\quad .
\label{tauc}
\end{equation}

CR is usually identified by a minimum (resp. a maximum) in the coefficient of variation 
$R$ (resp. in $\tau_c$) occurring at finite noise variance \cite{lindner_ojalvo_2004}.

\subsection{Results}

\begin{figure}[h]
\begin{center}
\includegraphics*[width=6cm]{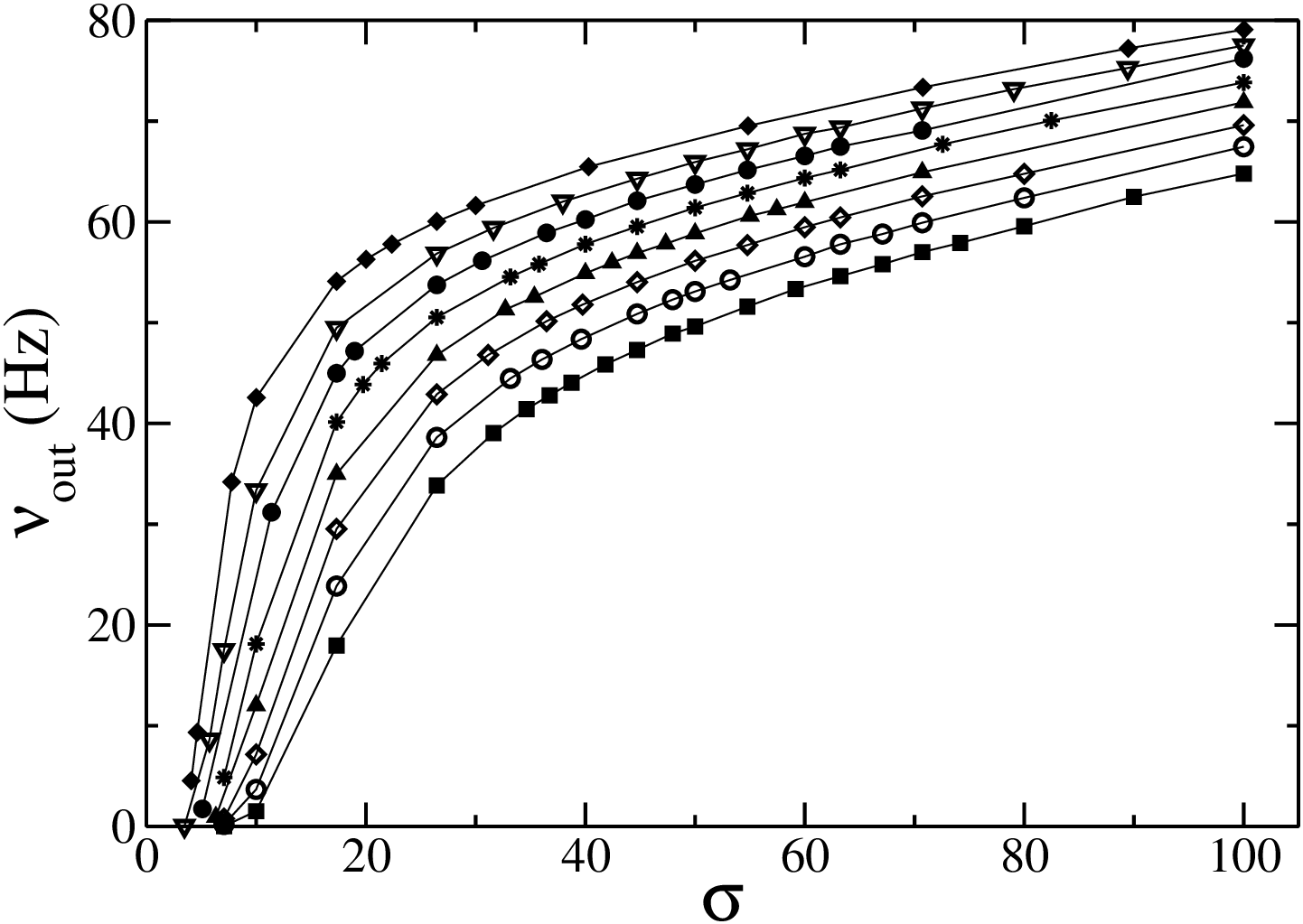}
\end{center}
\caption{Silent regime: average firing rate $\nu_{out}$ as 
a function of $\sigma$ for increasing values of $\bar I$ 
(from the bottom to the top $\bar I$=-1,0,1,2,3,4,5,6.15 $\mu$A/cm$^2$).}
\label{ivar_ave.total}
\end{figure}

The HH neuron subjected to stochastic input exhibits 
a noise-induced (irregular) spiking behaviour with an average firing rate $\nu_{out}=1/A_{ISI}$.
For fixed $\bar I$ the firing rate increases with 
the standard deviation of the noise (cf. Fig.~\ref{ivar_ave.total}). 

The noise-induced firing activity becomes 
more and more pronounced by approaching the repetetive firing bifurcation 
$I_{SN}$ for the deterministic HH. The HH model is a type II neuron, therefore
the onset of oscillation at $I \simeq I_{SN}$ is associated with a finite frequency
($\simeq 50$ Hz), however, in presence of noise, arbitrarily low spike rates can be observed
even in the silent regime.

\subsubsection{Response of the HH model in the low and high noise limit}
\label{response}

\begin{figure}[h]
\begin{center}
\includegraphics*[width=6cm]{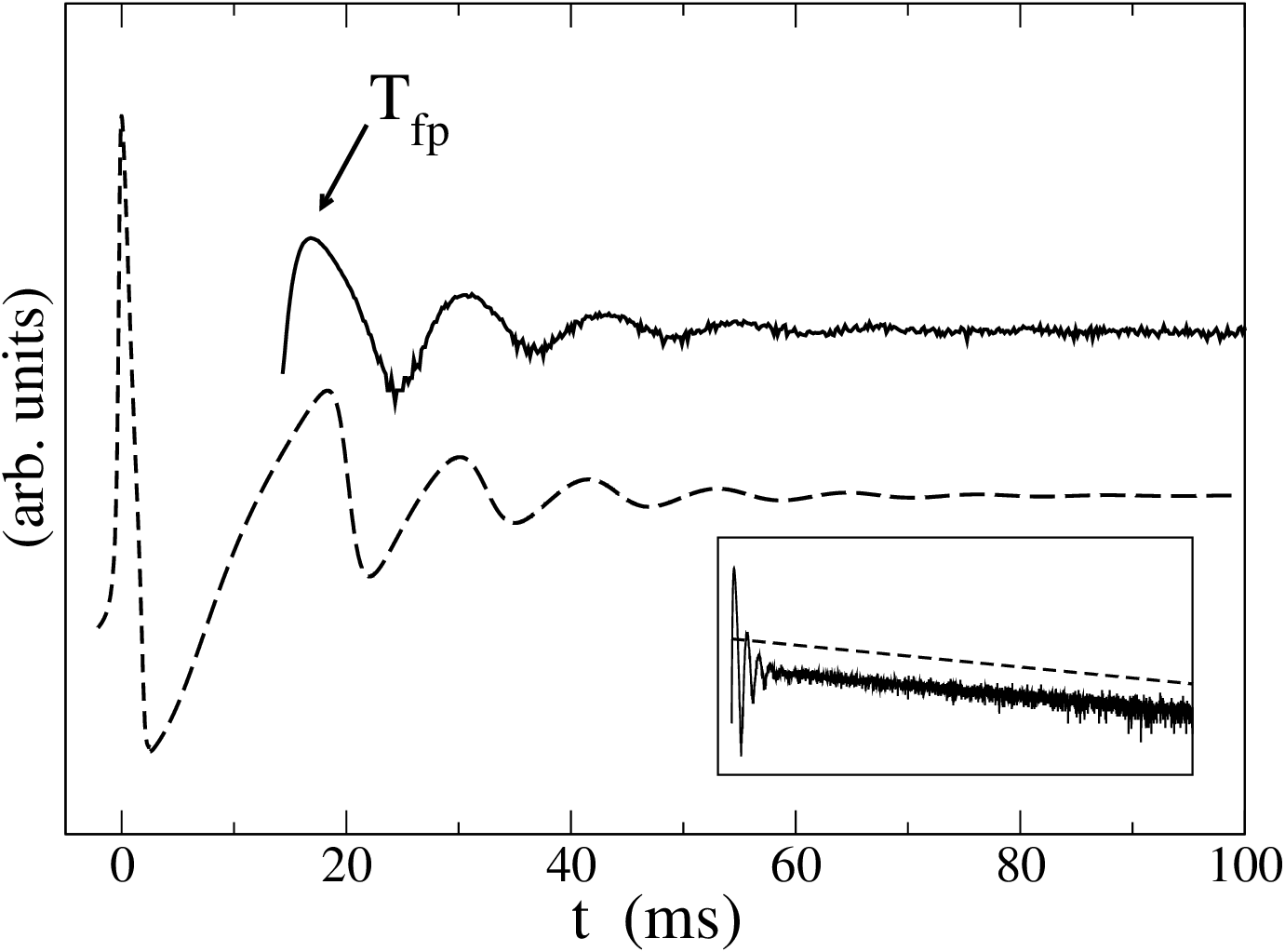}
\end{center}
\caption{$P_{ISI}(t)$ obtained for the stochastic input
(\ref{I_imp}) with $\sigma=4.6$ (continuous curve) and the
relaxation dynamics of the membrane potential $V(t)$ following a step 
current stimulation (dashed line). The
position of the output spike has been shifted to $t=0$ and
the action potential has been rescaled to better
reveal the relaxation oscillations. In the inset
the exponential tail due to the activation mechanism
is shown. The data refer to $\bar I =6.15 \mu$A/cm$^2$.
} \label{multi}
\end{figure}

Let us now examine more in detail the origin of noise-induced spiking
in the low and high noise limits.
For low noise we observe the coexistence of
multi-peaks and an exponential tail in the $P_{ISI}(t)$ (cf. Fig.~\ref{multi}). 
As explained in the following the multi-peak structure
is related to the relaxation dynamics of the membrane potential $V(t)$
following a spike emission, while the exponential tail is associated with noise 
induced activation processes.

To better understand the first mechanism we have studied the evolution of $V(t)$
following a step current of amplitude $\bar I$, i.e.,
\begin{equation}
I(t)=\left\{
\begin{array}{ll} 0  &
\mbox{if  $t \le 0$}     \\
\bar I
& \mbox{if $t>0$}
\end{array} \right.
\qquad .
\label{step}
\end{equation}
As shown in Fig.\ref{multi} (dashed line) $V(t)$ exhibits one spike followed by damped oscillations.
This suggests that the probability of eliciting a second spike is enhanced in
correspondence the maxima of the relaxation oscillations, where $V(t)$ is nearest to the firing threshold.
Moreover, it is possible to show that the period $T_{nl}$ of the first oscillation
has a clear nonlinear origin, while the period of successives oscillations $T_l$ can be explained by 
a linear stability analysis around the fixed point.
Indeed we have verified that the first peak ($T_{fp}$) observed in the distribution $P_{ISI}(t)$ 
(continuous curve in Fig.\ref{multi}) is related to $T_{nl}$,
while the successives peaks in the $P_{ISI}(t)$ correspond to the
linear oscillations of period $T_l$ \cite{noi}. 

The second firing mechanism responsible for the exponential tail in the $P_{ISI}(t)$ is due to the
competition between two effects, on one side the tendency of the dynamics to relax
towards its stable fixed point and on the other side noise fluctuations that instead lead
the system towards the firing threshold. 
In this framework the dynamics of $V(t)$ can be described as the overdamped dynamics
of a particle in a potential well under the influence of thermal
fluctuations \cite{Kramers} and the average firing time, or activation time $T_a$, can be expressed 
in terms of the Kramers expression \cite{Kramers} for sufficiently small noise:
\begin{equation}
T_a \propto {\rm e}^{W/\sigma^2}
\quad ,
\label{rate}
\end{equation}
where $\sigma^2$ plays the role of
an effective temperature and $W$ of an energy barrier.
For $\sigma^2 < W$ , i.e., in the low noise limit, the dynamics can be characterized as an activation
process: $P_{ISI}(t)$ resembles a Poissonian distribution with $R \approx 1$.
For $\sigma^2 > W$, i.e., in the high noise limit, the multi-peak structure and the exponential tail 
disappear and $P_{ISI}(t)$ reduces to an inverse Gaussian corresponding to
the distribution of the first passage times for a diffusive process plus drift~\cite{tuckwell}.

\begin{figure}[h]
\begin{center}
\includegraphics*[clip,width=6cm]{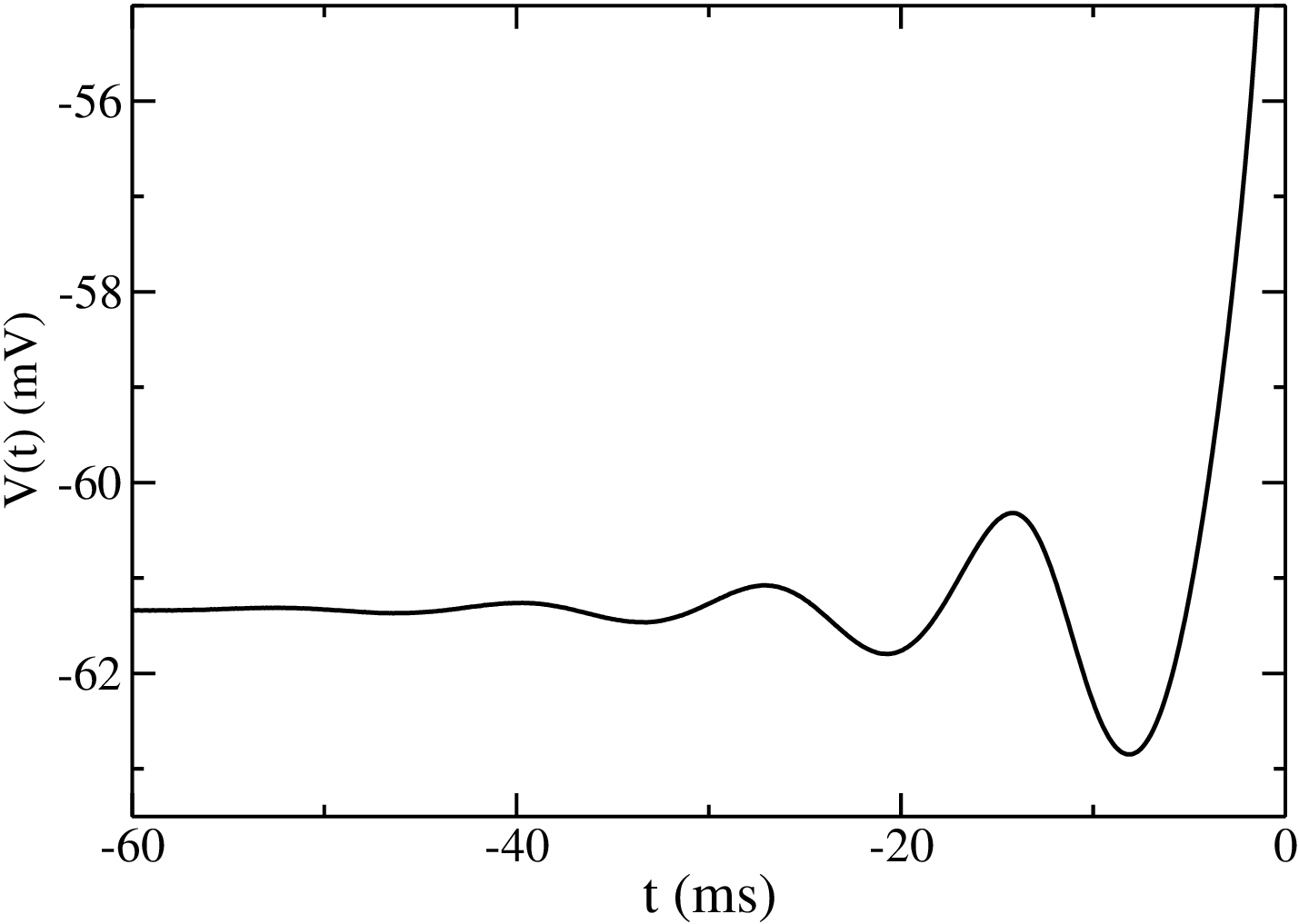}
\end{center}
\caption{$\bar I= 5 \mu A/cm^2$ and $\sigma=5.7$. STAP preceding the emission of a spike.  At time $t=0$
the potential overcomes the spike detection threshold (fixed at $-5$ mV).
STAP has been calculated by averaging only over ISIs longer
than $150$ ms.
}
\label{stap}
\end{figure}

As a further indicator we have estimated
the spike triggered average potential (STAP)~\cite{bialek}, that gives the average shape of $V(t)$ preceding the emission of a spike,
for sufficiently long ISIs and for small $\sigma$.
As shown in Fig.~\ref{stap}, $V(t)$ oscillates with period $\approx T_l$
before firing; therefore the emission of a spike (for long ISIs) is triggered by
the excitation of linear subthreshold oscillations around the fixed point.
Thus the HH neuron nearby the rest state acts as a sort of selective filter
since it responds (by emitting a spike) with higher probability when it is stimulated 
with a specific input frequency ($\approx 1/T_l=61-88$ Hz for $ 0 \le \bar I \le I_{SN}$).
This result agrees with the analysis of~\cite{yu_wang_2001}
where it has been shown that a silent HH neuron subjected
to a sinusoidal current optimally resonates when forced with a
frequency linearly correlated with $1/T_l$.

\subsubsection{Coherent response of the Hodgkin-Huxley neuron}
\label{tre2}

In the silent regime we have found the coexistence of two CRs: one is related to the regularization of
the emitted spike trains at intermediate noise levels and 
has previously been reported in~\cite{lee_neiman_1998,yu_wang_2001}; the
second one, observed for the first time, occurs at very low noise and is associated to the coherence 
of subthreshold oscillations.

\begin{figure}[h]
\begin{center}
\includegraphics*[width=8cm]{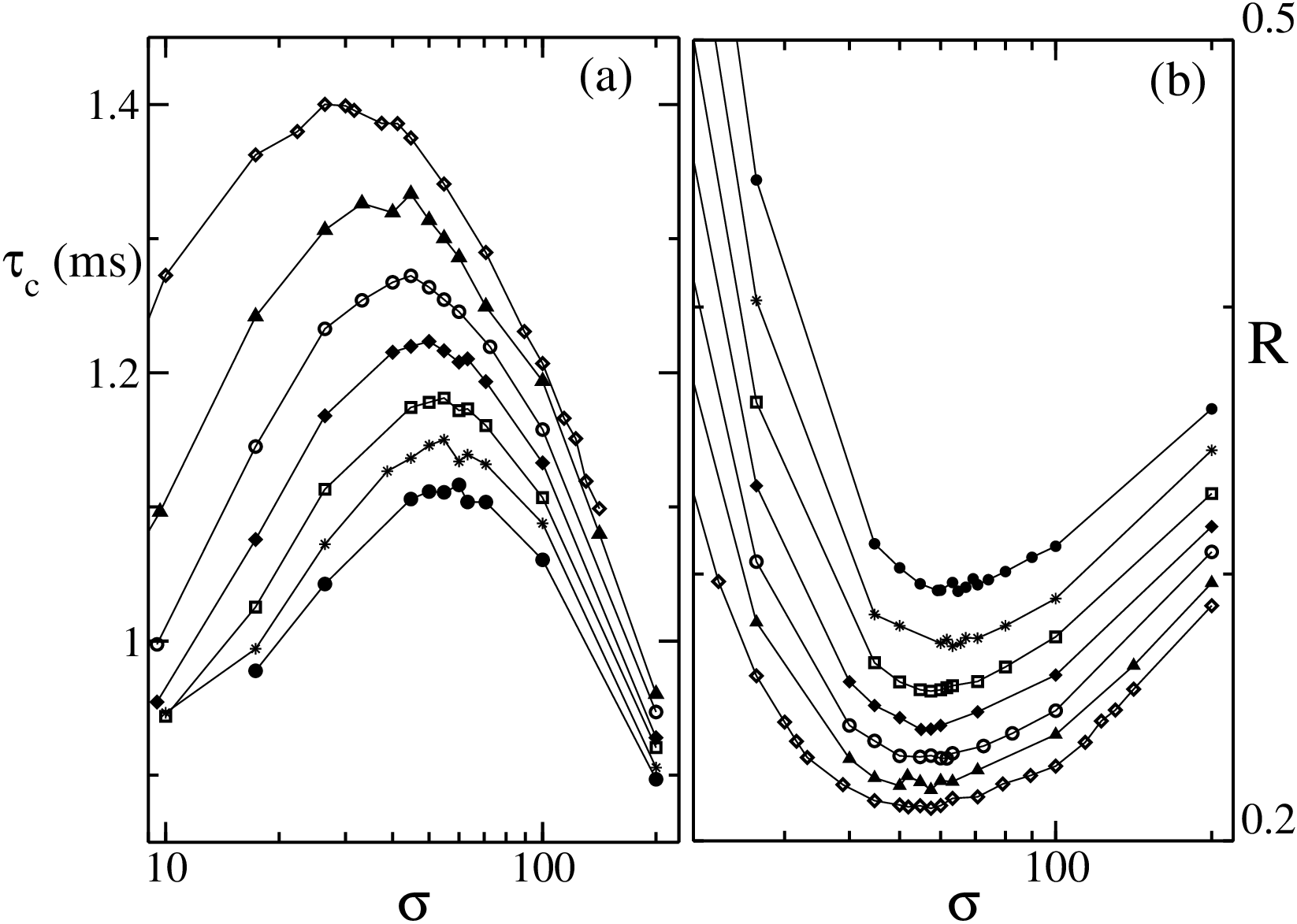}
\end{center}
\caption{Coherence of the emitted spike trains for increasing values of $\bar I$ in the silent regime. (a) $\tau_c$ 
(from bottom to top $\bar I$=-1,0,1,2,3,4,5 $\mu$A/cm$^2$). (b) $R$ (from top to 
bottom $\bar I$=-1,0,1,2,3,4,5 $\mu$A/cm$^2$ ).}
\label{ivar_tauc_cv}
\end{figure}

The first CR can be explained by noticing that the dynamics of excitable systems
can be characterized in terms of two characteristic times~\cite{pikovsky_kurths_1997}: 
an activation time $T_a$, representing 
the time needed to excite the system, and an excursion time $T_e$, 
indicating the duration of the spike (i.e., the time spent in the excited state).
An ISI is given by the sum of these two times and therefore $R$ can be (formally) expressed as the 
sum of two parts depending separately on $T_a$ and $T_e$. 
These two contributions vary in an opposite way when the noise is increased and
the competition of these two tendencies leads to the maximal coherence (associated to a minimum in $R$) 
for finite noise~\cite{pikovsky_kurths_1997,lindner_ojalvo_2004}.
So there is an initial range of $\sigma$-values where by increasing the noise the spike emission is facilitated 
($T_a$ reduces accordingly to Eq. (\ref{rate})) and this renders the response of the neuron more and more regular.
On the other side at high $\sigma$-values the response becomes again more irregular, because 
the noise induces firing even during the relative refractory period
thus modifying even the duration $T_e$ of the spike itself.
To summarize, the activation process 
responsible for firing is gradually
substituted by a diffusive process with drift and at the transition from one mechanism to the other
a regularization of the output signal is observed.
As shown in Fig.~\ref{ivar_tauc_cv}, for different values of $\bar I$, both the coefficient of variation $R$ and the signal correlation time $\tau_c$ 
were able to identify clearly the first kind of coherence in the silent regime.  
We can also see that $\tau_c$ increases and $R$ decreases when $\bar I \rightarrow I_{SN}$, i.e., the coherence of the 
emitted spike trains increases by approaching the repetetive firing regime.

\begin{figure}[h]
\begin{center}
\includegraphics*[width=4.2cm]{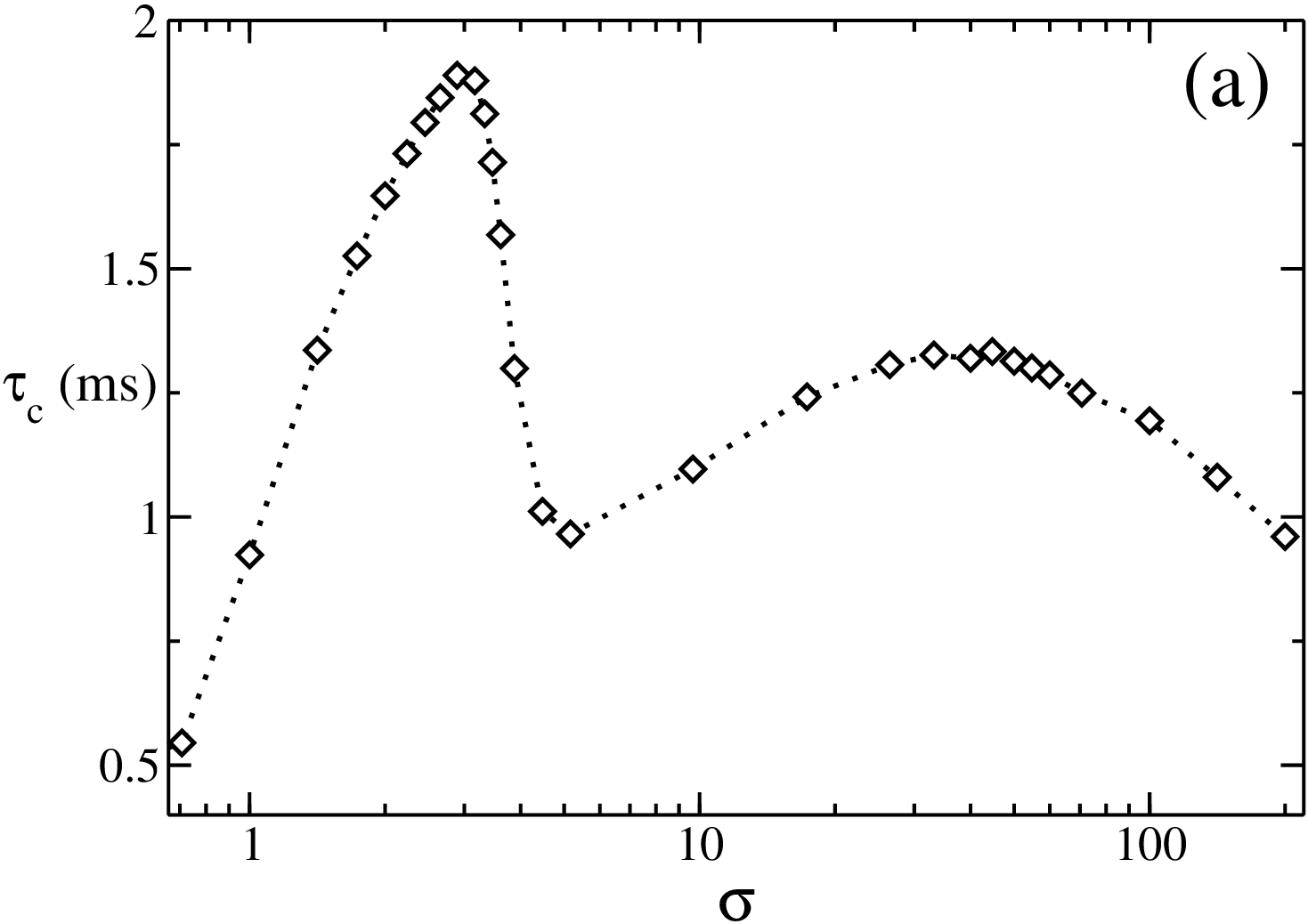}
\includegraphics*[width=4.2cm]{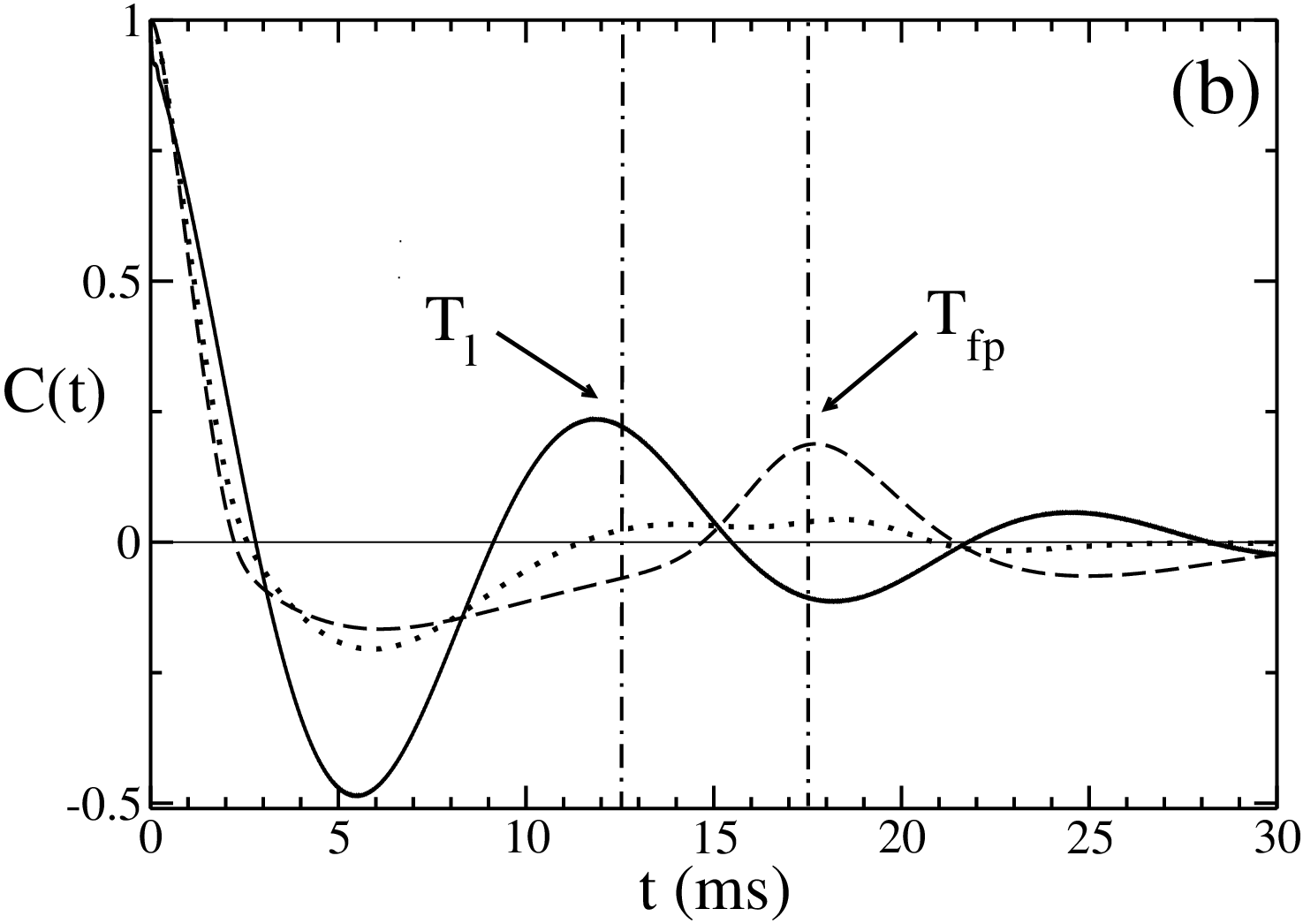}
\end{center}
\caption{$\bar I= 4 \mu A/cm^{2}$: (a) correlation time $\tau_c$ 
(empty circles) as a function of the standard deviation of 
the input noise $\sigma$: (b) autocorrelation function
$C(t)$ for three increasing noise values $\sigma=2.9$ (continuous line), 4.5 (dotted line) and 9.7 (dashed line); 
the vertical dot-dashed lines indicates the period $T_l$ and the period $T_{fp}$ for $\sigma=9.7$.}
\label{tauI4}
\end{figure}

The second kind of resonance could only be detected by $\tau_c$.
In fact this property is not related to spike emission (suprathreshold dynamics) but to the behaviour of the signal below the firing threshold.
In Fig.\ref{tauI4}a the behaviour of $\tau_c$ for $\bar I= 4 \mu A/cm^{2}$ is shown in a wider range of noise 
with respect to Fig.~\ref{ivar_tauc_cv}a. 
In this range $\tau_c$ exhibits indeed two maxima: the first and higher maximum 
at low noise ($\sigma \approx 3$) is related to the coherence of subthreshold oscillations, while the second one 
at intermediate noise ($\sigma \approx 33$) is due to the CR just discussed above.

 Let us now gain deeper insight about the origin of these two maxima.
 For extremely low noise ($\sigma < 3$) the neuron rarely fires, i.e., $T_a \rightarrow \infty$; increasing the noise the subthreshold oscillations generated 
 by the input kicks are more and more correlated.
 For $\sigma > 3$ the occurrence of rare spikes tends to decorrelate the signal leading to a
 decrease of $\tau_c$. 
 With a further increase of the noise $V(t)$ becomes essentially a sequence of spikes and 
 $C(t)$ represents the correlation between them; a second peak appears in $\tau_c$ indicating the 
maximal coherence of the spike sequence.

To investigate the transition from one to the other behaviour let us analyze the autocorrelation function of $V(t)$
for increasing value of $\sigma$, as shown in Fig.~\ref{tauI4}b.
For $\sigma \approx 3$ the autocorrelation function $C(t)$ oscillates
with a period $\approx T_l$ (see Sec.\ref{response}), while at $\sigma \approx 9.7$ the maxima of
$C(t)$ correspond to multiples of $T_{fp}$, i.e., the first peak of $P_{ISI}(t)$.
In between these two values there is a transition where the course of $V(t)$, initially consisting of 
mere subthreshold oscillations, is dominated by the spikes.
The transition is observed for $\sigma \approx 4.5$, where in correspondence to the first oscillation  $C(t)$ reveals 
two maxima, one located at $t \approx T_l$ and another at $t \approx T_{fp}$.

\subsection{Conclusions}

We have presented an analysis of the response of the silent HH model in the {\it high-input regime}. In this regime the HH neuron 
displays a large variety of dynamical behaviors and its response is completely determined by the average and the variance 
of the stochastic input.  Our main result is the coexistence of two different coherence resonances:
the first one, at intermediate noise levels, associated with the regularization of the spikes emitted by the neuron;
the second one, at very low noise, related to the coherence of subthreshold oscillations around the fixed point.
The first one can be revealed using both the ISI coefficent of variation $R$ and the autocorrelation time 
of the signal $\tau_c$. The second type of resonance, observed for the first time, is not related to
spike emission and can thus only be detected by means of $\tau_c$.

Moreover, we have examined the behaviour of the HH neuron for low and high input noise variance. 
The richness of the HH dynamics is 
particularly pronounced for low input noise where we have found the coexistence of
a multimodal structure and an exponential tail in the $P_{ISI}(t)$.
The peaks in the $P_{ISI}(t)$ suggests that the system,
under the influence of stochastic inputs, can resonate when forced 
with specific frequencies: the first peak is associated to frequencies in the $\gamma$-range~\cite{sheperd}
(namely, from $40$ to $66$ Hz for $\bar I \in [0:9] \enskip \mu A/cm^2$),
and the second one to lower frequencies (namely, from $30$ to $37$ Hz
for the same interval of currents) \cite{noi}.

In the silent regime the responsiveness of the single neuron is enhanced by
stochastic stimulations, in particular the regularity of the emitted
spike trains is maximal in correspondence to an optimal
noise level. Moreover, we expect that a population of such neuronal elements will have the
capability to exhibit coherent and correlated activity over
different time scales (mainly in the $\gamma$ and $\beta$-ranges
~\cite{sheperd}), a property that is believed to be important for
information encoding for cortical
processing~\cite{salinas_seinoski_nature}. Indeed it has been
found ~\cite{yu_wang_2001} that for sufficiently
strong synaptic coupling a globally coupled HH
network subjected to stochastic inputs reveals a collective synchronized rhythmic
firing in a range of $40-60$ Hz, induced via CR.

To proceed in the direction of more realistic situations the present analysis performed for 
a current-driven model should be extended by considering conductance-based synaptic 
inputs~\cite{tiesinga_2000,rudolph_2003}. It is surely worth to address this issue in forthcoming
studies, because the response of these two classes of models to noise fluctuations can sometimes be
even opposite, as shown in~\cite{tiesinga_2000} for the output firing rate.

%%%%%%%%%%%%%%%%%%%%%%%%%%%%%%%%%%%%%%%%%%%%%%%%%%%%%%%%%%%%%%%%%%%%%%%%%%%%%%
%       References
%%%%%%%%%%%%%%%%%%%%%%%%%%%%%%%%%%%%%%%%%%%%%%%%%%%%%%%%%%%%%%%%%%%%%%%%%%%%%%

\end{document}